 \documentclass[final,5p,times,twocolumn,authoryear,linenumbers]{elsarticle}

\usepackage[colorlinks=true,linktocpage=true,linkcolor=blue,citecolor=magenta,urlcolor=blue]{hyperref}
\usepackage[T1]{fontenc}
\usepackage{newtxtext}
\usepackage{amsmath,amssymb}
\usepackage{aas_macros}
\usepackage{graphicx}
\usepackage{xcolor}
\usepackage{bm}
\usepackage{lipsum}
\usepackage{academicons}
\definecolor{orcidlogocol}{HTML}{A6CE39}

\newcommand{\YL}{Y_{L_e}}
\newcommand{\fzero}{f_{0}}

\journal{Physics Letters B}

\begin{document}

\begin{frontmatter}



\title{Radial Modes of Hot Neutrino-Trapped Hybrid Stars}


\author[label1]{Peter B. Rau} 
\affiliation[label1]{organization={Columbia Astrophysics Laboratory, Columbia University},
            city={New York},
            postcode={10027},
            state={NY},
            country={USA}}

\author[label2]{Andrea Sabatucci} 

\affiliation[label2]{organization={Institute of Theoretical Physics, University of Wroc\l aw},
            city={Wroc\l aw},
            postcode={50-204},
            country={Poland}}

\author[label2,label3]{Armen Sedrakian} 

\affiliation[label3]{organization={Frankfurt Institute for Advanced Studies},
            city={Frankfurt am Main},
            postcode={60438},
            country={Germany}}

\begin{abstract}
 We study the fundamental radial oscillation mode of hot hadronic and
hybrid stars under conditions relevant to proto-neutron stars and
neutron-star merger remnants.  The stellar backgrounds are constructed
from finite-temperature equations of state with a covariant density
functional for hadronic matter and three-flavor 2SC quark matter at
high density. We consider two isentropic configurations in both neutrino-transparent matter and with
trapped electron neutrinos at fixed electron-lepton fraction. For trapped matter we compare
a Gibbs-type construction with global electron-lepton-number
conservation to a locally constrained Maxwell-type prescription; in
both cases the fixed-entropy stellar equation of state contains an
extended coexistence region and no sharp interface requiring
conversion rate-dependent junction conditions.
We find that the fundamental radial frequency follows the corresponding
purely hadronic sequence up to the onset of deconfinement and then
decreases relative to it.  Neutrino trapping shifts this dynamical
signature to larger stellar masses and central pressures, reflecting
the delayed appearance of quark matter.  The Maxwell-type prescription
moves the onset of deconfinement to still higher central pressure and
produces a more rapid decrease of the mode frequency near the stability
limit.  We quantify the deconfinement effect through the fractional
frequency shift between hybrid and hadronic stars at equal
gravitational mass.  These results show that the fundamental radial mode
provides a sensitive dynamical probe of the interplay among entropy,
lepton trapping, phase-transition construction, and quark
deconfinement in hot compact stars.
\end{abstract}




\end{frontmatter}




\section{Introduction}
\label{sec:intro}

Whether quark degrees of freedom are realized in the interiors of compact
stars is a major unresolved issue in the study of strongly interacting
matter at supranuclear density.  Mature neutron stars predominantly
constrain cold, neutrino-transparent matter close to weak equilibrium,
with temperatures well below the characteristic Fermi scales.  The
conditions encountered in proto-neutron stars formed after core collapse
and in binary neutron-star merger remnants are markedly different
(for reviews, see~\citet{Lattimer2016,Foucart2023,Sedrakian2023}).
Shock and compression heating can raise the temperature to several tens
of MeV, while the large neutrino opacity delays deleptonization and the
relaxation toward neutrino-free beta equilibrium
\cite{Prakash1997,Pons1999,Espino2024,Pajkos2025}.  The thermodynamic
state must therefore be specified by thermal and compositional variables
in addition to the baryon density, for example, by the entropy per
baryon and the total lepton fraction in neutrino-opaque matter.  Through
their influence on the chemical potentials and phase-equilibrium
conditions, finite entropy and trapped leptons can shift the onset of
deconfinement, modify the extent of the coexistence region, and change
the strength of the hadron--quark transition along astrophysically
relevant trajectories.

Radial oscillations provide a particularly direct measure of the
dynamical response of a spherical stellar configuration.  The lowest,
nodeless radial mode determines stability against spherically
symmetric perturbations, while its frequency is sensitive to the
compressibility of the stellar interior.  Radial pulsations of hot
proto-neutron stars were studied in ~\citet{Gondek1997},
where finite entropy and neutrino trapping were shown to modify the
mode spectrum.  More recently,~\citet{Sun2021} investigated radial
oscillations of cold hybrid stars containing a Gibbs-constructed
hadron--quark mixed phase and found that the fundamental mode vanishes
at the maximum-mass configuration, consistent with the usual
turning-point stability criterion.  The effects of entropy and fixed
lepton fraction on the radial modes of isentropic proto-neutron stars
without quark phase were subsequently studied in~\citet{Sun2025},
which showed that both thermal effects and neutrino trapping can
substantially modify the mode frequencies and their evolution toward
the cold neutron-star limit.

In hybrid stars with a sharp first-order phase boundary, the radial-mode problem acquires an additional dependence on the conversion rate between the two phases.  If phase conversion is slow or rapid compared with the oscillation period, different interface junction conditions apply and the stability boundary need not coincide in the two limits \cite{Haensel1989,Pereira2018,Goncalves2022,Rau2023,Rau2023b}.

The present work studies radial modes of hot hybrid stars with an extended hadron--quark coexistence region, allowing us to isolate how neutrino trapping shifts the dynamical signature of deconfinement.
For that purpose we use the finite-temperature hybrid equations of state developed in~\citet{Sabatucci2026}, in which the phase transition along the relevant isentropic trajectories is represented by an extended coexistence region rather than by a sharp density discontinuity. In neutrino-trapped matter, local electric neutrality and global conservation of electron-lepton number lead to a Gibbs-type mixed phase whose pressure varies continuously with density.  The same framework provides continuous isentropic hybrid equations of state for the neutrino-transparent sequences considered below.  Consequently, there is no isolated hadron--quark interface at which the slow- and rapid-conversion junction conditions of sharp-interface models must be imposed.

Our purpose is to determine how this equilibrium mixed phase manifests itself in the fundamental radial mode of hot stars. We compare purely hadronic and hybrid configurations at fixed entropy per baryon, and with and without neutrino trapping. The central questions are whether the onset of deconfinement produces an identifiable change in the radial-mode spectrum and how that change is displaced by neutrino trapping.  We additionally introduce a simple dimensionless measure of the dynamical effect of deconfinement, namely the fractional difference between the fundamental frequencies of hybrid and hadronic configurations at the same gravitational mass.

The paper is organized as follows.  Section~\ref{sec:eos} summarizes the stellar equations of state and equilibrium backgrounds.  Section~\ref{sec:modes} reviews the relativistic radial-mode equations and specifies the treatment of the mixed phase.  Our numerical results are presented in Sec.~\ref{sec:results}, and Sec.~\ref{sec:conclusions} summarizes the main conclusions.

\section{Hot hadronic and hybrid-star backgrounds}
\label{sec:eos}

We employ the same finite-temperature equations of state as in~\citet{Sabatucci2026} and therefore give only the information required for the oscillation analysis.  The hadronic sector is described by the DDLS family of covariant density functionals.  In the present work we use the reference DDLS50 parametrization, with symmetry-energy slope $L_{\rm sym}=50$ MeV and $Q_{\rm sat}=400$ MeV~\cite{Tsiopelas2024EPJA}.  The high-density phase is modeled within a three-flavor Nambu--Jona-Lasinio description with repulsive vector interactions and two-flavor color superconductivity (2SC).  We use the reference quark-matter parametrization of~\citet{Sabatucci2026}, with $\eta_V=G_V/G_S=1$ and $G_D/G_S=1$.  The phenomenological pressure shift that fixes the effective location of deconfinement is also kept at its reference value.

The hybrid equation of state is constructed at finite temperature along isentropic trajectories.  We consider four thermodynamic conditions,
\begin{eqnarray}
(S,\YL)&=& (1,0.4),\ (2,0.4),\quad {\rm trapped}~\nu_e
\label{eq:cases_a}\\
(S,Y_{\nu}) &=& (1,0),\ (2,0),\quad {\rm untrapped}~\nu_e
\label{eq:cases_b}
\end{eqnarray}
where $S=1$ or $2$ is the entropy per baryon in units of $k_{\rm B}$, $Y_{\nu}= (n_{\nu_e}-n_{\bar{\nu}_e})/{n_B} = 0$ denotes the neutrino-transparent sequence and $\YL=0.4$ the electron-neutrino-trapped sequence.  For each thermodynamic condition we construct both a purely hadronic sequence and a hybrid sequence.  This pairing is essential below because it allows the dynamical effect of deconfinement to be separated from the changes caused by finite entropy and lepton trapping themselves.

For trapped neutrinos, baryon number and electron-lepton number are conserved globally during phase coexistence, while each phase is locally electrically neutral.  The quark volume fraction $\chi$ changes continuously through the mixed region, and the pressure is a continuous function of the density.  As shown in~\citet{Sabatucci2026}, trapped neutrinos stabilize the hadronic phase against deconfinement and shift the onset of the mixed phase toward higher density.  The main purpose of the present work is to determine how this thermodynamic delay is reflected in the radial-mode spectrum.

In neutrino-transparent matter, where only baryon number remains as an independent globally conserved charge, the phase transition is described by a Maxwell-type construction. For trapped matter, we additionally construct a locally constrained Maxwell-type prescription for comparison with the globally constrained Gibbs-type construction.
For neutrino-transparent matter, phase equilibrium at fixed temperature
requires
\begin{equation*}
 P_H=P_Q,\qquad T_H=T_Q,\qquad \mu_B^H=\mu_B^Q .
\end{equation*}
For neutrino-trapped matter, the electron-lepton fraction $\YL$ is fixed
to the same value in each phase rather than being redistributed globally.
The corresponding chemical-equilibrium condition is equality of the
Gibbs free energy per baryon,
\begin{equation*}
 \mu_B^H+\YL\,\mu_{L_e}^H
 =
 \mu_B^Q+\YL\,\mu_{L_e}^Q ,
\end{equation*}
together with equality of the temperature and pressure.

At fixed temperature, the Maxwell-type construction has the usual
density discontinuity at the coexistence pressure.  The stellar
equations of state considered here, however, are isentropic.  The
isentropes of the two pure phases intersect the phase boundary at
different points, and the fixed-$S$ trajectory between these points is
constructed as a two-phase coexistence region.  At each point along the
phase boundary the quark volume fraction $\chi$ is determined from
\begin{align*}
 n_B &= (1-\chi)n_H+\chi n_Q,\\
 S n_B &= (1-\chi)S_H n_H+\chi S_Q n_Q ,
\end{align*}
where $n_H$ and $n_Q$ are the baryon densities and $S_H$ and $S_Q$ the
entropy per baryon in the two phases.  The energy density is then
\begin{equation*}
 \epsilon=(1-\chi)\epsilon_H+\chi\epsilon_Q ,
\end{equation*}
while the pressure satisfies $P=P_H=P_Q$ at each point on the
coexistence line.  Because the temperature and chemical potentials vary
along that line, the pressure is not constant along the isentropic
trajectory, and both $n_B$ and $\epsilon$ vary continuously.  Thus the
final isentropic Maxwell-based equation of state contains an extended
coexistence interval rather than a sharp density discontinuity.

At low density, the finite-entropy equations of state are matched at
$n_B=0.08~{\rm fm}^{-3}$ to the corresponding DDLS low-density equation
of state at fixed entropy and without trapped neutrinos.  This matching
is not strictly thermodynamically consistent.  Since the same
low-density prescription is used for each matched hadronic--hybrid pair,
its effect is expected to be largely common to the two configurations,
but it remains a limitation of the present treatment.

The stellar backgrounds are obtained by solving the Tolman--Oppenheimer--Volkoff equations for each isentropic equation of state.  We work in units $G=c=1$ and write the metric as
\begin{equation}
 ds^2=-e^{2\nu(r)}dt^2+e^{2\lambda(r)}dr^2+r^2d\Omega^2.
\end{equation}
The equilibrium equations are
\begin{align}
 \frac{dP}{dr}&=-\frac{(\epsilon+P)(m+4\pi r^3P)}{r^2(1-2m/r)},\\
 \frac{dm}{dr}&=4\pi r^2\epsilon,
\end{align}
where $\epsilon$ is the energy density.  Integration from a chosen central pressure $P_c$ to $P(R)=0$ determines the gravitational mass $M=m(R)$ and radius $R$.

\begin{figure}[t]
\centering
\includegraphics[width=\linewidth]{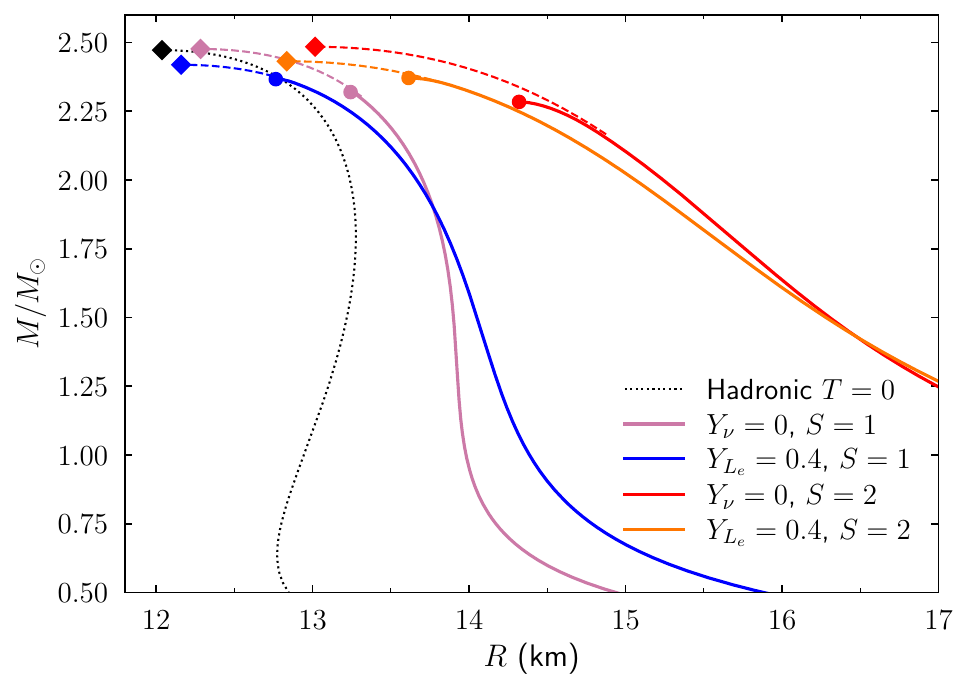}
\caption{Gravitational mass as a function of stellar radius for the hadronic and hybrid (Gibbs construction) equations of state used in the radial-mode calculation. The four thermodynamic conditions correspond to $S=1,2$ with neutrino-transparent matter ($Y_{\nu} = 0$) and trapped electron neutrinos at $\YL=0.4$. The purely hadronic counterparts of hybrid configurations are shown with the same color dashed lines. The dotted black curve corresponds to hadronic equation of state at $T=0$ in $\beta$-equilibrium. Filled circles (diamonds) on stellar sequences mark the maximum-mass configuration for the mixed phase (purely hadronic) models.
}
\label{fig:MRbackground}
\end{figure}
The resulting mass--radius relations are shown in Fig.~\ref{fig:MRbackground}.
For each thermodynamic condition we display both the hybrid sequence
and its purely hadronic counterpart.  The hybrid sequences remain
compatible with maximum masses above $2\,M_\odot$, while finite entropy
and neutrino trapping produce appreciable shifts in the stellar radii
and in the location of the maximum-mass configuration.  These
background sequences provide the equilibrium stellar models used in
the radial-mode calculations below.

\section{Fundamental radial mode}
\label{sec:modes}

We compute linear, adiabatic radial oscillations on the static backgrounds described above.  We use the formulation employed previously in Refs.~\cite{Gondek1997,Rau2023,Rau2023b}.  The perturbation variables are the dimensionless radial displacement $\xi=\xi_{\rm dim}/r$ and the Lagrangian pressure perturbation $\Delta P$.  For a mode with angular frequency $\omega$, they obey
\begin{equation}
\frac{d\xi}{dr}=\left(\frac{d\nu}{dr}-\frac{3}{r}\right)\xi
-\frac{\Delta P}{r\Gamma P},
\label{eq:xi}
\end{equation}
and
\begin{align}
\frac{d\Delta P}{dr}={}&
\left[
 e^{2\lambda}\left(\omega^2e^{-2\nu}-8\pi P\right)
 +\frac{d\nu}{dr}\left(\frac{4}{r}+\frac{d\nu}{dr}\right)
\right]
(\epsilon+P)r\xi \nonumber\\
&-\left[
\frac{d\nu}{dr}+4\pi(\epsilon+P)re^{2\lambda}
\right]\Delta P.
\label{eq:dP}
\end{align}
Since we consider adiabatic perturbations, the relativistic adiabatic
index is written as
\begin{equation}
 \Gamma=\frac{\epsilon+P}{P}
 \left(\frac{dP}{d\epsilon}\right)_{\rm eq},
\label{eq:Gamma}
\end{equation}
where the derivative is taken along the same equilibrium isentropic
equation of state that is used to construct the stellar background.
For neutrino-trapped matter this corresponds to a trajectory at fixed
entropy per baryon $S$ and fixed electron-lepton fraction $\YL$,
whereas for neutrino-transparent matter the corresponding
beta-equilibrated isentrope is used.  Thus, the particle fractions
and, within the coexistence region, the quark volume fraction are
allowed to adjust according to the equilibrium conditions as the
energy density changes.

In practice, Eq.~\eqref{eq:Gamma} is evaluated by differentiating the
one-parameter relation $P(\epsilon)$ along the appropriate equilibrium
trajectory.  The present calculation therefore adopts the equilibrium
adiabatic response of the mixed phase.  A frozen-composition or
finite-relaxation treatment~\cite{Ghosh2025}, in which the particle fractions and/or
the phase fraction cannot readjust on the oscillation timescale, would
in general lead to a different adiabatic index and is not considered
here.
  
Regularity at the stellar center gives
\begin{equation}
 \Delta P(0)=-3\Gamma P\,\xi(0),
\end{equation}
and we choose the arbitrary normalization $\xi(0)=1$.  At the stellar surface the Lagrangian pressure perturbation vanishes,
\begin{equation}
 \Delta P(R)=0.
\end{equation}
Equations~(\ref{eq:xi}) and (\ref{eq:dP}) constitute an eigenvalue problem for $\omega^2$.  We determine the eigenvalues by a shooting method and identify the nodeless solution as the fundamental radial mode, $\fzero=\omega_0/2\pi.$ A configuration is dynamically stable against radial perturbations as long as $\omega_0^2>0$.

A crucial distinction from the sharp-interface problem studied in Refs.~\cite{Haensel1989,Pereira2018,Rau2023,Rau2023b} is that the present hybrid equations of state contain a continuous equilibrium coexistence region.  There is therefore no isolated radius at which a density discontinuity requires separate slow- or rapid-conversion junction conditions for $\xi$ and $\Delta P$.  Instead, Eqs.~(\ref{eq:xi}) and (\ref{eq:dP}) are integrated continuously through the mixed phase using the local equilibrium equation of state and its adiabatic response.  The same statement applies to the isentropic Maxwell-based sequences used below for comparison.  Although the underlying isothermal Maxwell construction contains a density discontinuity, the fixed-entropy trajectory traverses an extended coexistence interval in which the temperature and phase fraction vary continuously.  The resulting $P(\epsilon)$ is therefore continuous, and no separate interface junction condition is required in the radial-mode calculation.

This treatment assumes that the equilibrium mixed-phase description is appropriate for the radial perturbation problem.  Finite relaxation times associated with changes of the phase fraction, weak equilibration, or lepton redistribution could in principle modify the adiabatic response of the coexistence region.  Such nonequilibrium effects are distinct from the sharp-interface slow/rapid conversion problem and are beyond the scope of the present calculation.

\section{Results}
\label{sec:results}

\subsection{Fundamental frequency along the stellar sequences}

\begin{figure}[t]
  \centering
  \includegraphics[width=\linewidth]{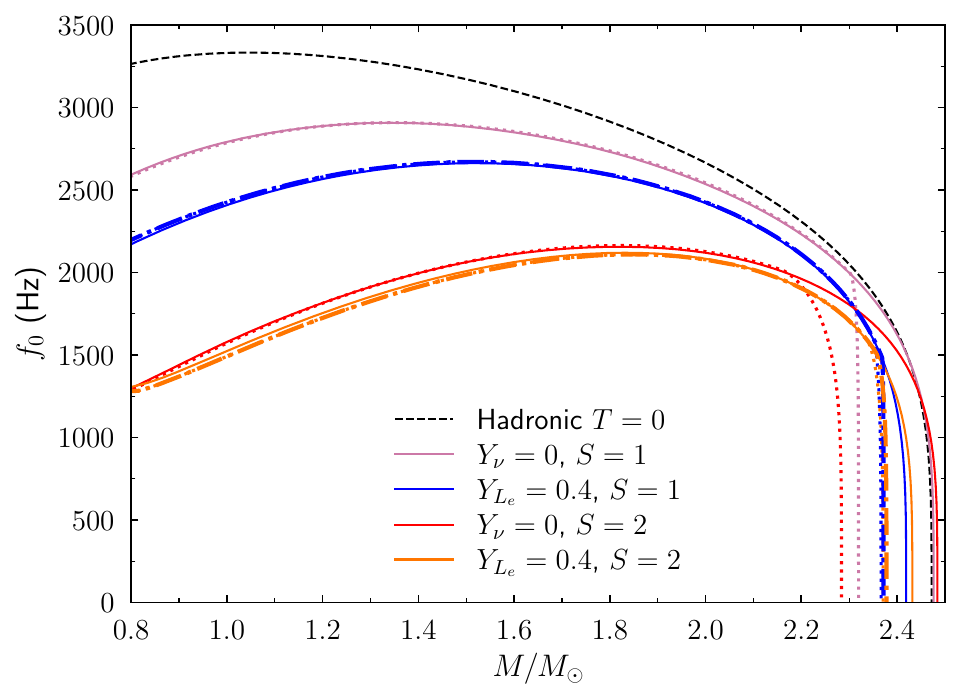}
  \caption{
Fundamental radial-mode frequency $f_0$ as a function of gravitational
mass for purely hadronic stars (solid lines), the baseline hybrid
sequences with an extended isentropic coexistence region (dotted lines),
and, for trapped matter, hybrid stars obtained with the locally
constrained Maxwell-type prescription (dot-dashed lines).  The baseline
hybrid sequences are shown for $S=1$ and $2$, both with trapped electron
neutrinos at $\YL=0.4$ and in neutrino-transparent matter with
$Y_\nu=0$.  The hybrid curves coincide with their hadronic counterparts
before the onset of deconfinement and soften after quark-containing
matter appears.
  }
\label{fig:fM}
\end{figure}

Figure~\ref{fig:fM} shows the fundamental radial frequency as a function of gravitational mass for the four thermodynamic conditions in Eqs.~\eqref{eq:cases_a} and \eqref{eq:cases_b}. The corresponding purely hadronic sequences are plotted together with the hybrid sequences so that the effect of quark deconfinement can be read directly at fixed mass.  For the trapped-neutrino cases of Eq.~\eqref{eq:cases_a}, we additionally show sequences obtained with the locally constrained Maxwell-type prescription described in Sec.~\ref{sec:eos}.  Although the corresponding isothermal Maxwell transition contains a density jump, the fixed-entropy stellar equation of state contains an extended coexistence interval and is continuous in both pressure and energy density.  The radial perturbation equations are therefore integrated continuously through this region, without imposing a sharp-interface conversion condition.

At masses below the onset of the mixed phase, the hadronic and hybrid calculations describe the same physical stellar core and their frequencies coincide, up to numerical accuracy.  Once the central density enters the coexistence region, the hybrid sequence begins to depart from the corresponding hadronic one.  The fundamental frequency is reduced relative to the purely hadronic value and falls increasingly rapidly as the maximum-mass configuration is approached.  Thus the appearance of quark-containing matter produces a direct dynamical counterpart of the softening associated with the phase transition.

The absolute frequency scale depends strongly on the thermodynamic state.  Increasing the entropy from $S=1$ to $S=2$ changes both the stellar background and the radial restoring force, producing a substantial shift of the mode frequencies.  Neutrino trapping has a second, qualitatively different effect: because it delays deconfinement, the separation between the hadronic and hybrid curves occurs at a larger stellar mass.  The location of the radial-mode signature therefore tracks the trapping-induced displacement of the phase transition found from the thermodynamic analysis of~\citet{Sabatucci2026}.

\begin{figure}[t]
  \centering
  \includegraphics[width=\linewidth]{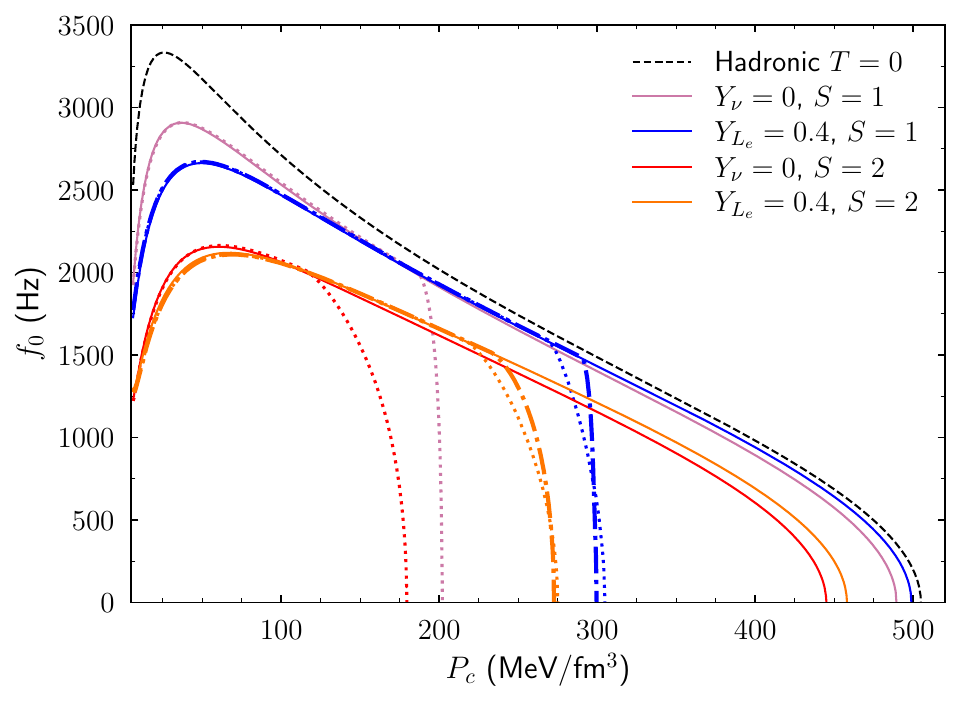}
\caption{Fundamental radial-mode frequency $f_0$ as a function of central pressure for the same stellar sequences as in Fig.~\ref{fig:fM}. The central-pressure representation emphasizes where the hybrid-mode spectrum begins to depart from the corresponding purely hadronic sequence.}
\label{fig:fPc}
\end{figure}

The same information is shown as a function of central pressure $P_c$ in Fig.~\ref{fig:fPc}.  This representation makes the approach to the radial stability boundary especially transparent.  For each sequence, $f_0$ rises from the low-central-pressure configurations, reaches a broad maximum, and subsequently decreases toward zero as the limiting configuration is approached.  In the hybrid models, entry into the mixed phase modifies this high-density decrease and leads to an earlier departure from the corresponding hadronic-mode curve.

The central-pressure representation also shows directly that the trapping-induced delay of deconfinement is a high-density effect rather than a uniform rescaling of the mode spectrum.  At lower $P_c$, where the stellar center remains hadronic, the paired curves are nearly indistinguishable.  Their difference develops only after the pressure increases enough to allow for the coexistence region.  For the trapped sequences, the locally constrained Maxwell-type prescription shifts the onset of quark-containing matter to a higher central pressure relative to the globally constrained Gibbs-type construction.  The mode frequency subsequently decreases more rapidly and reaches zero at a slightly lower central pressure.

\subsection{Fractional mode softening due to deconfinement}

To isolate the dynamical effect of the phase transition from the overall thermal and leptonic shifts of the mode spectrum, we compare the hybrid and purely hadronic frequencies at the same gravitational mass.  We define
\begin{equation}
 \Delta_f(M)
 =\frac{\fzero^{\rm had}(M)-\fzero^{\rm hyb}(M)}
 {\fzero^{\rm had}(M)}.
\label{eq:ratio}
\end{equation}
With this convention, $\Delta_f>0$ corresponds to a softening of the fundamental radial mode induced by the appearance of quark-containing matter.

Below the deconfinement threshold, Eq.~(\ref{eq:ratio}) vanishes
because the two stellar models have the same hadronic core.  It
becomes nonzero once the hybrid sequence develops a mixed-phase region
and then increases toward the high-mass end of the sequence.  This
quantity therefore provides a compact measure of the asteroseismic
imprint of deconfinement while largely factoring out the common shift
caused by entropy and neutrino trapping.
\begin{figure}[t]
  \centering
  \includegraphics[width=\linewidth]{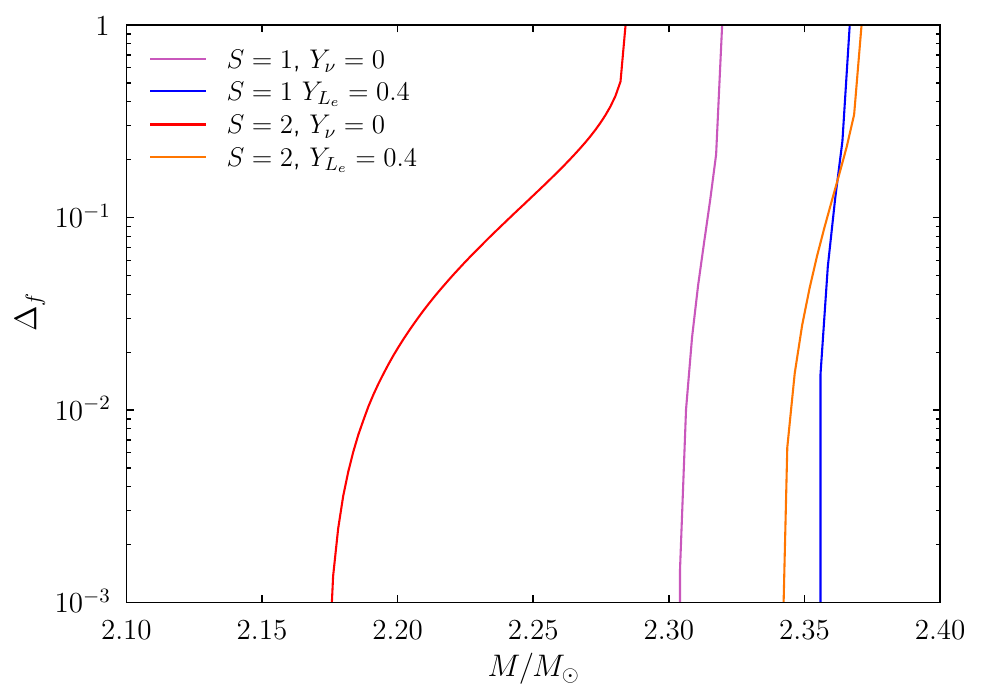}
\caption{Fractional reduction of the fundamental radial-mode frequency
  caused by deconfinement, defined by Eq.~\eqref{eq:ratio},
  for thermodynamic configurations described in Eqs.~\eqref{eq:cases_a}--\eqref{eq:cases_b} with a mixed phase.}
\label{fig:ratio}
\end{figure}
Figure~\ref{fig:ratio} shows
$\Delta_f(M)$ for the four thermodynamic conditions considered here.
Neutrino trapping displaces the rise of $\Delta_f$ to larger
gravitational mass, directly reflecting the delayed phase transition,
while the $S=2$ neutrino-transparent sequence exhibits the broadest
mass interval over which a sizeable softening develops.

Table~\ref{tab:HybridSequences} lists key properties of
the hybrid stellar models' mass and pressure. It is seen that a
softening exceeding 10\% is concentrated in the final $\simeq 0.005-
0.043 M_{\odot}$ of the stable sequence, depending on entropy and
lepton content.  For the present quark-matter parameter set
this mass interval is relatively narrow.  However, variations of the
quark-matter model parameters are known to modify the extent of the
mixed phase and may therefore lead to a broader stellar-mass range over
which a sizeable mode softening occurs.

\begin{table}[tb]
\renewcommand{\arraystretch}{1.2}
    \centering
    \begin{tabular}{ccccc}
    \hline\hline
     & (1,0) & (1,0.4) & (2,0) & (2,0.4)\\
     \hline 
    $P_{\rm mp}$ & 187 & 270 & 116 & 220 \\
    $M(P_c=P_{\rm mp})$ & 2.303 & 2.338 & 2.164 & 2.342 \\
    $M_{\rm max}$ & 2.320 & 2.367 & 2.284 & 2.371 \\
    $P_c(f_0=0)$ & 202 & 305 & 179 & 275 \\
    $\Delta M(\Delta_f\geq0.1)$ & 0.0053 & 0.0062 & 0.0426 & 0.0127 \\
    \hline\hline
    \end{tabular}
    \caption{Properties of the hybrid stellar sequences: pressure
      $P_{\rm mp}$ at which the mixed phase starts, lowest stellar
      mass containing the mixed phase $M(P_c=P_{\rm mp})$, maximum
      stellar mass $M_{\rm max}$, central pressure at which $f_0$
      vanishes $P_c(f_0=0)$, and the range of stellar masses for which
      $\Delta_f\geq0.1$, $\Delta M(\Delta_f\geq0.1)$.  The models are
      denoted by their values of $(S,Y_\nu)$ or $(S,\YL)$,
      following
      Eqs.~\eqref{eq:cases_a}--\eqref{eq:cases_b}: the
      first and third columns, $(1,0)$ and $(2,0)$, correspond to
      neutrino-transparent matter with $Y_\nu=0$, whereas the second
      and fourth columns, $(1,0.4)$ and $(2,0.4)$, correspond to
      neutrino-trapped matter with fixed electron-lepton fraction
      $\YL=0.4$.  }
    \label{tab:HybridSequences}
\end{table}

\subsection{Stability of mixed phase hybrid stars}

The fundamental radial mode provides the direct criterion for dynamical stability against spherically symmetric perturbations: stable configurations satisfy $f_0^2>0$, whereas the onset of instability occurs at $f_0^2=0$.  For the equilibrium mixed-phase response considered here, we find that the zero of the fundamental mode coincides, within numerical accuracy, with the maximum-mass turning point of the corresponding stellar sequence, in accordance with the Bardeen--Thorne--Meltzer (BTM) criterion \cite{Thorne1966,Bardeen1966}.  This behavior should be contrasted with hybrid stars containing a genuine sharp first-order interface, for which nonequilibrium conversion physics can alter the junction conditions and permit stability beyond an ordinary turning point~\cite{Rau2023,Rau2023b}.  In the present equilibrium coexistence construction, no such sharp-interface degree of freedom is present.  The relevant future nonequilibrium question is instead whether the phase fraction and composition can readjust on a timescale shorter than the radial oscillation period.

The Maxwell-type sequences provide an additional consistency check.  At fixed temperature, the underlying Maxwell construction contains the usual density discontinuity at the coexistence pressure.  This discontinuity is absent, however, from the fixed-entropy stellar equation of state.  Along an isentrope the system traverses the coexistence line while the phase fraction changes continuously from zero to unity; the temperature and pressure therefore vary across the coexistence interval, and both the baryon and energy densities remain continuous.  The radial perturbation equations can consequently be integrated continuously through this region without imposing a sharp-interface conversion condition.  Using the same equilibrium adiabatic response as for the stellar background, we find that the fundamental mode again vanishes at the maximum-mass turning point, in accordance with the BTM criterion.  Thus the slow/rapid-conversion ambiguity familiar from sharp-interface models is not operative for the isentropic Maxwell-type sequences considered here.

\section{Discussion and conclusions}
\label{sec:conclusions}

We have calculated the fundamental radial oscillation mode of hot hadronic and 2SC hybrid stars for isentropic matter with $S=1$ and $2$, both in the neutrino-transparent regime and with trapped electron neutrinos at $\YL=0.4$.  The equations of state are those developed in~\citet{Sabatucci2026}, allowing the dynamical consequences of the same finite-temperature phase-transition physics to be studied without introducing an additional phenomenological stellar model.

The first result is that deconfinement leaves a visible imprint on the fundamental radial mode even when the phase transition is represented by an extended coexistence region.  The hybrid and hadronic mode frequencies coincide while the stellar center remains in the hadronic phase.  Once the mixed phase appears, the hybrid frequency softens relative to the purely hadronic value, and the difference grows toward the high-mass end of the stellar sequence.  The fractional frequency shift in Eq.~(\ref{eq:ratio}) provides a direct quantitative measure of this effect.

The second result is that neutrino trapping displaces the mode frequency softening in hybrid stars toward larger mass and central pressure.  This behavior has a transparent thermodynamic origin.  A fixed electron-lepton fraction increases the electron and proton abundances, stabilizing the hadronic phase against quark deconfinement and shifting the transition to higher density~\cite{Sabatucci2026}.  The radial spectrum therefore carries a dynamical record of the same delayed transition: the hadronic and hybrid frequencies remain close over a larger portion of the stellar sequence before separating.

Entropy changes the absolute mode frequencies substantially, but it does not remove this basic pattern.  The comparison of $S=1$ and $S=2$ sequences shows that thermal structure and deconfinement affect the mode spectrum in distinguishable ways: entropy changes the overall frequency scale, whereas the appearance of the mixed phase produces a differential softening relative to a hadronic star under the same thermodynamic conditions.  This motivates the use of $\Delta_f(M)$ rather than the absolute value of $f_0$ alone as a diagnostic of the phase transition.

For trapped matter, the comparison between the globally constrained Gibbs-type construction and the locally constrained Maxwell-type prescription shows that the treatment of lepton-number conservation also affects the location of the dynamical signature.  The Maxwell-type prescription shifts the onset of quark-containing matter to higher central pressure and produces a more rapid decrease of the fundamental frequency near the stability limit.  Nevertheless, because the stellar trajectories are isentropic, both constructions contain an extended coexistence interval and lead to continuous equilibrium equations of state.

The present calculation also clarifies the relation to earlier studies of radial stability in sharp-interface hybrid stars.  Because the isentropic coexistence regions used here are continuous, no slow, rapid or intermediate (reactive)-conversion junction condition is required.  This eliminates the interface ambiguity that can qualitatively alter radial stability when a density discontinuity is present~\cite{Rau2023,Rau2023b}.  It does not, however, imply that all conversion dynamics are irrelevant.  If the phase fraction, weak composition, or lepton content cannot follow a millisecond radial oscillation adiabatically, the effective adiabatic index of the mixed phase may differ from the equilibrium one used here. Extending the present mixed-phase calculation to a frozen-composition or finite-relaxation response is an interesting problem for future work.

Radial modes of a nonrotating spherical star do not radiate gravitational waves. Rather,
their significance is that they characterize the dynamical response and stability timescale of hot compact stars and expose a sensitivity to deconfinement.  In realistic proto-neutron stars or merger remnants, rotation, nonspherical dynamics, and mode coupling can connect radial-like compression dynamics to observable fluid motion.  Determining whether the phase-transition-induced shifts found here survive in such environments will require multidimensional calculations.

In summary, the fundamental radial mode provides a simple dynamical diagnostic of phase transition
to hot quark matter, i.e., finite temperature/entropy deconfinement in proto-neutron stars and binary mergers.  The appearance of a 2SC-containing mixed phase softens the mode relative to a purely hadronic star, while neutrino trapping delays the onset of this softening in mass and central pressure. The effect is naturally quantified by a fractional hadronic--hybrid frequency difference at fixed mass.  These results establish a direct connection between the thermodynamic delay of deconfinement in lepton-rich matter and the radial dynamics of hot hybrid stars.

\section*{Acknowledgements}
P. B. Rau was supported by the Simons Foundation through a SCEECS postdoctoral fellowship (grant No. PG013106-02). A. Sabatucci and A. Sedrakian were supported by the Polish National Science Centre (NCN) Grant 2023/51/B/ST9/02798. 

\bibliographystyle{apsrev4-2}
\bibliography{2sc_eos}

\end{document}